\documentclass{pic2012}
\def\runauthor{Bhawna Gomber}
\def\shorttitle{}
\def\met{\ensuremath{\not\!\!{E_{T}}}}
\begin{document}

\title{Search for Dark Matter and Large Extra Dimensions in pp Collisions \\
Yielding a Photon and Missing Transverse Energy
}

\author{Bhawna Gomber}

\address{Saha Institute of Nuclear Physics, \\
1/AF Bidhannagar, Calcutta-700064, India\\
E-mail: bhawna.gomber@cern.ch }

\maketitle

\abstracts{Results are presented from a search for new physics in the final state containing a photon and missing transverse energy. The data corresponds to
an integrated luminosity of 5.0 fb$^{-1}$ collected in pp collisions at $\sqrt{s}$= 7 TeV by the CMS experiment. The observed event yield agrees with standard
model expectations. Using models for the production of dark matter particles($\chi$), we set 90$\%$ confidence level (C.L.) upper limits of 13.6-15.4 fb on χ
production in the $\gamma$+ $\met$ state. These provide the most sensitive upper limits for spin-dependent $\chi$-nucleon scattering for χ masses between 1 and
100 GeV. For spin dependent contributions, the present limits are extended to M($\chi$) $<$ 3.5 GeV. For ADD models with 3-6 large extra dimensions, our data
exclude extra-dimensional Planck scales between 1.64 and 1.73 TeV at 95$\%$ C.L.
}

\section{Introduction}
Proton-proton collisions at the Large Hadron Collider (LHC) resulting in final states consisting of a photon ($\gamma$) and missing transverse energy
($\met$), are used to investigate dark matter (DM) production and models of extra spatial dimensions (see further). 

 Dark matter is the dominant non-baryonic component of the universe's matter density~\cite{DMGeneral}. Searches for a DM candidate ($\chi$) in the universe takes the form of direct detection of ($\chi$) 
through elastic ($\chi$)-nucleon scattering or indirect detection via characterstic radiation from $\chi$ $\bar{\chi}$ annihilation in astrophysical sources.	
Since DM travels through the Compact Muon Solenoid detector~\cite{CMS} (CMS) without interaction, it appears as $\met$ in the resulting collision fragments.
At the LHC, DM can be produced in the reaction $q\bar{q}$ $\rightarrow$ $\gamma$ $\chi$ $\bar{\chi}$, where the photon is radiated by one of the incoming quarks.
The final state is a high-p$_{T}$ photon and $\met$. Recent theoretical work~\cite{DM1,DM4} casts this process in terms of a massive mediator in the $s$ channel that couples to a $\chi\overline{\chi}$ pair of Dirac particles.
This process is contracted into an effective theory with a contact interaction scale $\Lambda$, given by $\Lambda^{-2}$ = $g_{\chi}g_qM_M^{-2}$, where $M_M$ is the mediator mass and $g_{\chi}$ and $g_{q}$ are its couplings to $\chi$ and quarks, respectively.
The model provides a way to connect the $t$-channel $\chi$-nucleon elastic scattering to the $s$-channel pair-production mechanism.
The effective $s$-channel operator can be chosen to represent either a vector or axial-vector, spin-independent or spin-dependent interaction, respectively.

    The Arkani-Hamed, Dimopoulos, and Dvali (ADD) model~\cite{ADD} proposed a framework for
addressing the hierarchy problem, namely the existence of two widely
different fundamental scales of nature: the electroweak scale
($M_{EW}  \sim  10^3 GeV$) at which the electromagnetic and  weak
interaction unify, and the Plank scale ($M_{Pl} \sim 1.2 \times 10^{19} GeV$) at which
gravity becomes as strong as the gauge interactions. In this framework, space-time is postulated to have $n$ extra compact spatial dimensions with a characteristic scale $R$, leading to a modified Planck scale, $M_{D}$, given by $M_\mathrm{Pl}^2 \approx M_{D}^{n+2}R^n$.
Assuming $M_{D}$ is of the same order as $M_\mathrm{EW}$, the observed large value of $M_\mathrm{Pl}$ can be interpreted as being a consequence of the ``large" size of $R$ (relative to the Planck length $\approx M_\mathrm{Pl}^{-1}$) and the number of extra dimensions in the theory.
The process $q\bar{q}$ $\rightarrow$ $\gamma$G, where the graviton G escapes detection, motivates the search for events with single high-p$_{T}$ isolated photons.

\section{Event Selection}
Events are selected from a data sample corresponding to an integrated luminosity of 5.0 fb$^{-1}$ 
using single-photon triggers which are almost 100 $\%$ efficient within the selected signal region of $|\eta^\gamma|$ $<$ 1.44 and p$_{T}^{\gamma}$ $>$ 145 GeV.
Photons are required to pass additional isolation criteria. In particular, the scalar sum of p$_{T}$ depositions in the ECAL within a hollow cone of $0.06<\Delta R<0.40$, 
excluding depositions within $|\Delta\eta | = 0.04$ of the cluster center, must be $<$ 4.2 GeV + 0.006 $\times$  p$_{T}^{\gamma}$, the sum of scalar p$_{T}$ depositions in the HCAL within a
 hollow cone of $0.15<\Delta R<0.40$ must be $<$ 2.2 GeV + 0.0025 $\times$ p$_{T}^{\gamma}$, and the scalar sum p$_{T}$ of track  values in a hollow cone of $0.04<\Delta R<0.40$, excluding 
depositions that are closer to the cluster center than $|\Delta\eta|=$ 0.015, must be $<$ 2.0 GeV + 0.001 $\times$ p$_{T}^{\gamma}$ (with p$_{T}$ in GeV units).
The ratio of energy deposited in the HCAL to that in the ECAL within a cone of $\Delta R =$ 0.15 is required to be less than 0.05.
Due to the high luminosity at the LHC, there may be a multiple collisions at once which may result in an ambiguity in the assignment of a photon's origin collision vertex.
To account for this, the photon is required to pass the track isolation criteria for all reconstructed vertices and a systematic uncertainity is included.
It is also required that there are no pattern of hits in the pixel detector, called pixel seed. 

The $\met$ is defined by the magnitude of the vector sum of the transverse energies of all of the reconstructed objects in the event, and is computed using a particle-flow algorithm~\cite{PF}. The candidate events are required to have $\ensuremath{\not\!\!\!{E_{T}}}$ $>$ 130 GeV.
Events with excessive hadronic activity are removed by requiring that there are no particle flow jets, identified with the anti-k$_{T}$ algorithm with a distance parameter of 0.5~\cite{antikt} within 
$|\gamma|$ $<$ 3.0 outside of $\Delta R$(jet,$\gamma$) = 0.04 with p$_{T} >$ 20 GeV. To remove instrumental background arising from showers induced by bremsstrahlung from muons in the 
beam halo or in cosmic rays, it is required that the time assigned to the photon is $\pm$ 3 ns of that expected of a collision particle's arrival at the ECAL. Other anomalous ECAL signals are
reduced by requiring that the timing of all the deposits comprising the photon are consistent.
 
  After applying all of the selection criteria, 73 candidate events are found.

\begin{figure}[ht]
\begin{center}
\includegraphics[angle=0,totalheight=.31\textheight,width=.65\textwidth]{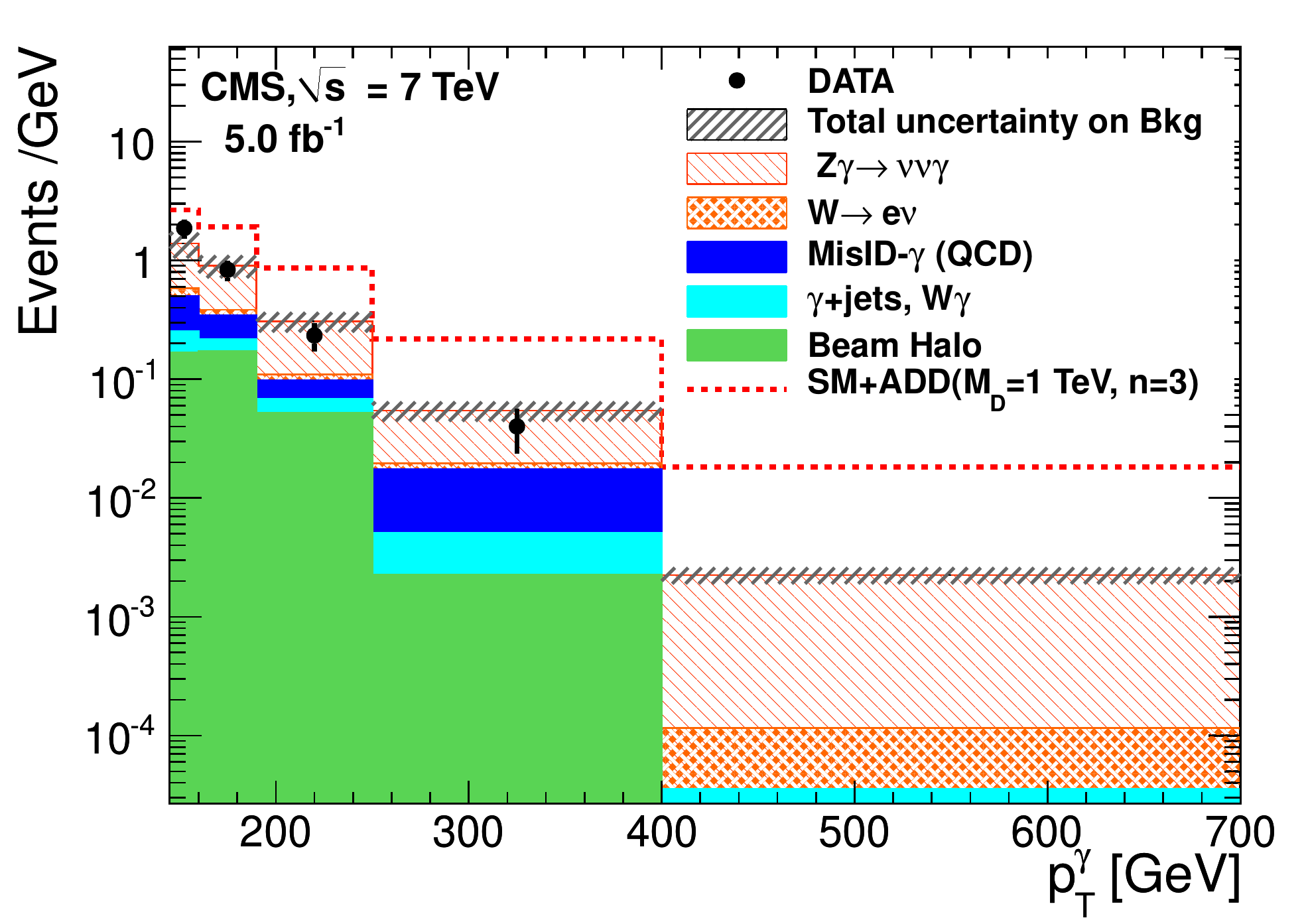}
\caption{The photon p$_{T}$ distribution for the candidate sample, compared with the estimated contributions from SM backgrounds and a prediction from ADD for $M_{D}=1 TeV$ and $n$~=~3.}
\label{fig:stack_plot}
\end{center}
\end{figure}

\begin{figure}[ht]
%\vspace*{7.0cm} 
\begin{center}
%\special{psfile=SI-limits-rec.ps voffset=-60 vscale=40
%hscale= 40 hoffset=10 angle=0}
%\special{psfile=SD-limits-rec-few.ps voffset=-60 vscale=40
%hscale= 40 hoffset=10 angle=0}
\includegraphics[angle=0,totalheight=.24\textheight,width=.45\textwidth]{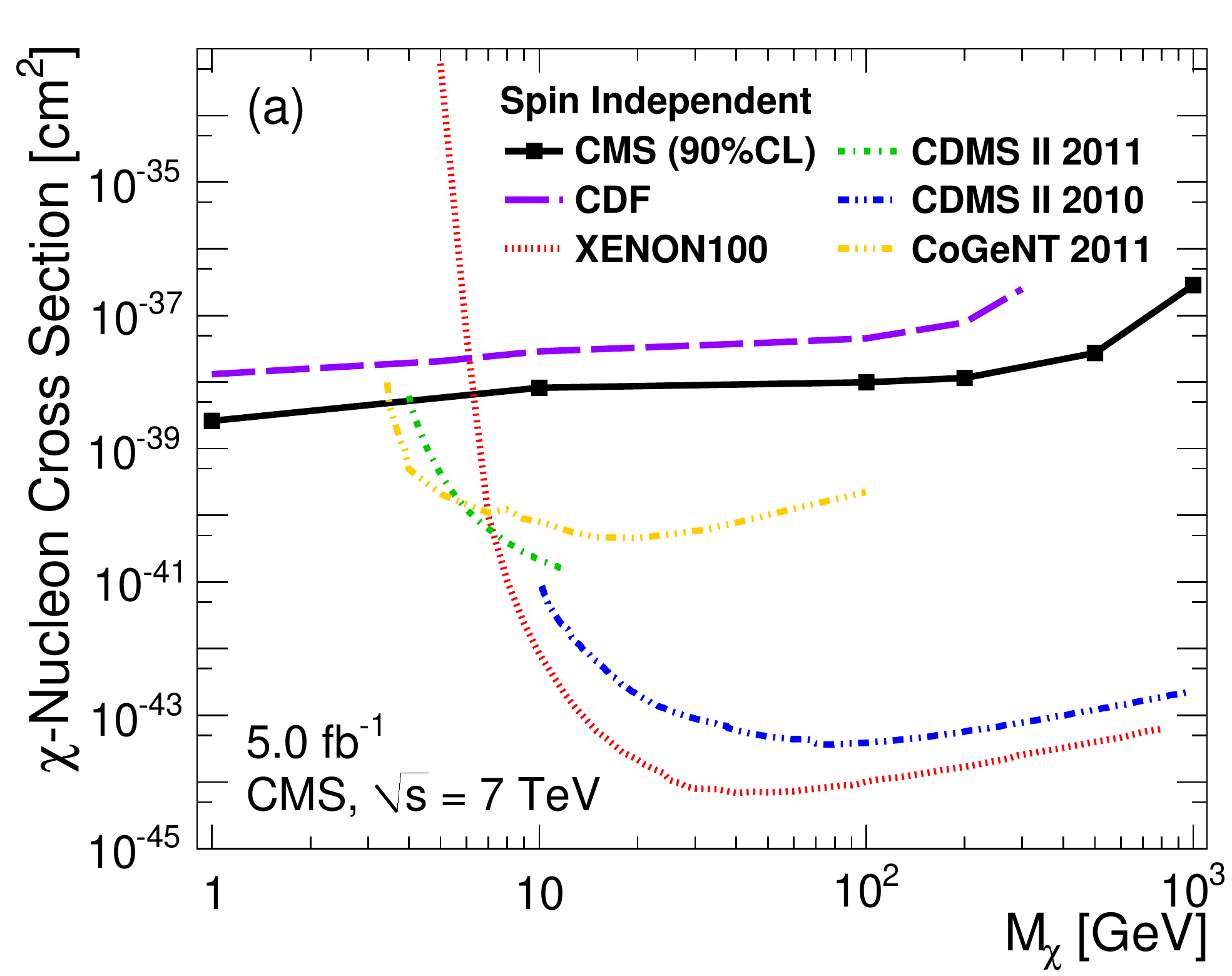}
\includegraphics[angle=0,totalheight=.24\textheight,width=.45\textwidth]{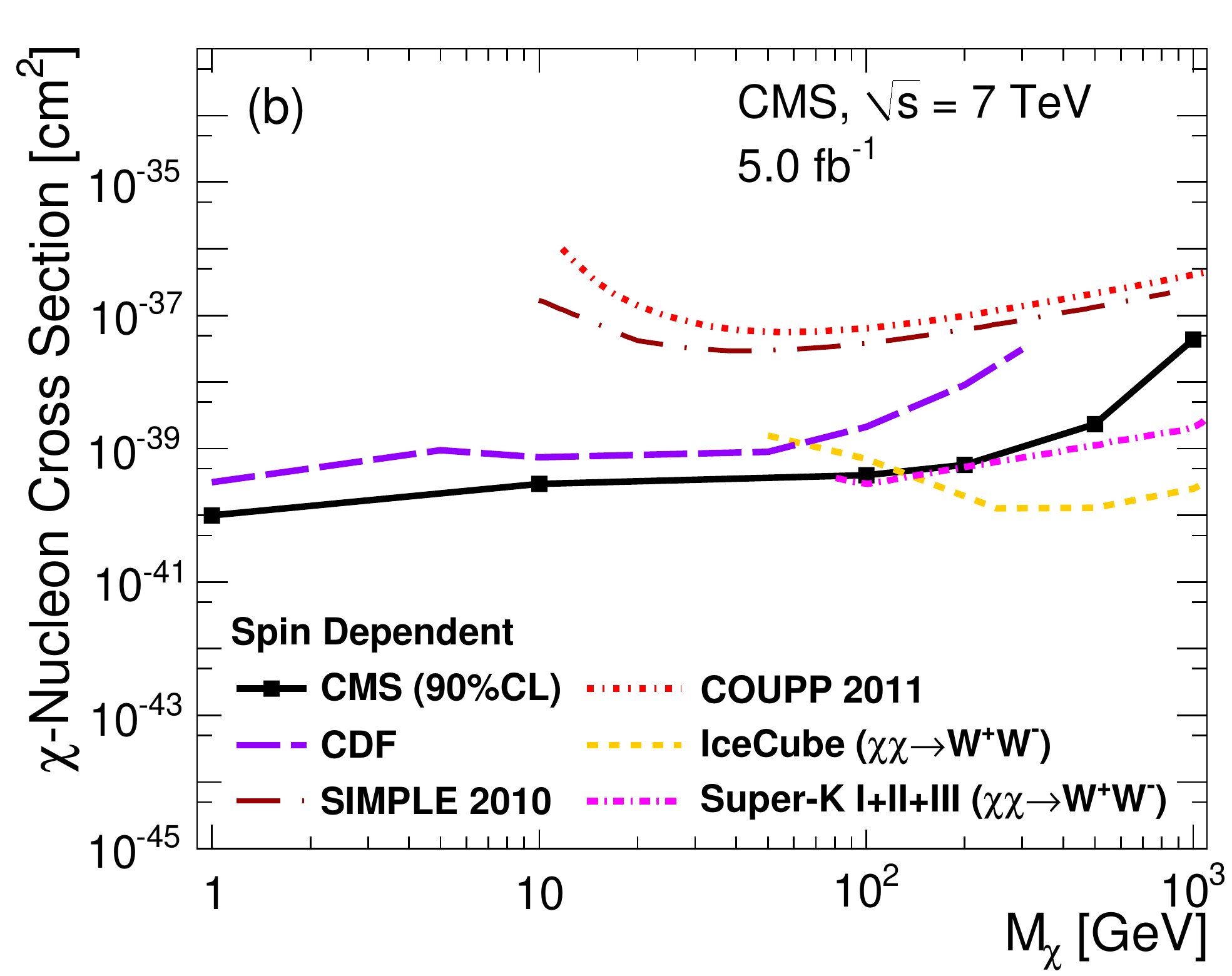}
\caption{The 90$\%$ CL upper limits on the $\chi$-nucleon cross section as a function of $M_{\chi}$ for (a) spin-independent and (b) spin-dependent scattering} 
 \label{fig:DMLimits}
%\centerline{\epsfxsize=2.9in\epsfbox{kim_mephi_lep.ps}}
\end{center}
\end{figure}

\begin{figure}[ht]
%\vspace*{7.0cm} 
\begin{center}
\includegraphics[angle=0,totalheight=.31\textheight,width=.65\textwidth]{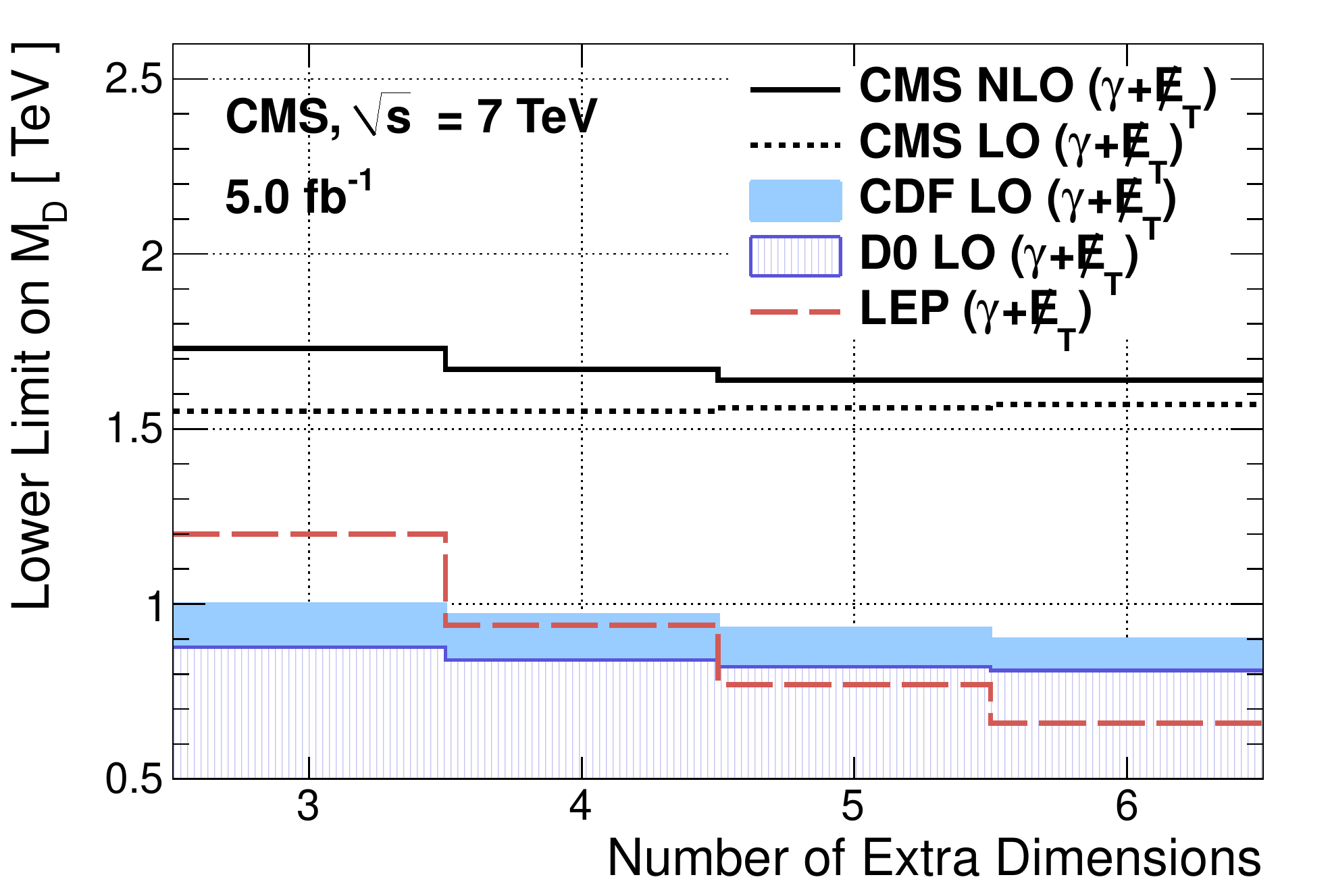}
\caption{Limits on $M_{D}$ as a function of n, compared to LO results from similar searches at the Tevatron and LEP}
\label{fig:ADDLimits}
%\centerline{\epsfxsize=2.9in\epsfbox{kim_mephi_lep.ps}}
\end{center}
\end{figure}

\section{Backgrounds}
The primary background for the $\gamma$ + $\met$ signal is the irreducible SM background from $Z\gamma\rightarrow\nu\bar{\nu}\gamma$ production. This and other SM backgrounds, including $W\gamma$, W$\rightarrow$e$\nu$, $\gamma$+jet, multijet (referred to as QCD), and diphoton events are estimated using Monte Carlo simulation.
The expected contribution from the $Z\gamma\rightarrow\nu\bar{\nu}\gamma$ process to the background is $45.3\pm 6.8$ events. The combined expected background from W$\gamma$, $\gamma$+jet and diphoton events is $4.1 \pm 1.0$.
      
      Backgrounds that are out of time with the collisions, are estimated from data by examining the transverse distribution of energy in the EM cluster and the time-of-arrival of the signal into the crystal with the largest energy deposition.
Templates for anomalous signals~\cite{spikes}, cosmic-ray muons, and beam halo events are fitted to a candidate sample that has no timing requirement. It is found that only 
halo muons gives a significant residual contribution to the in-time sample with an estimated $11.1\pm5.6$ events.
    
     Electrons misidentified as photon if its ECAL supercluster doesn't have a reconstructed pixel match. The matching of electron showers to pixel seeds has an efficiency of $\epsilon=0.9940\pm0.0025$, as estimated with Monte-Carlo(MC) simulated events and verified with \textit{Z}$\rightarrow$\textit{ee} events in data.
The contribution of \textit{W}$\rightarrow$\textit{e}\textit{$\nu$} events to the candidate sample has been found by scaling a control sample of electron candidates by $(1-\epsilon)/\epsilon$ yielding $3.5\pm1.5$ events.

     The contamination from jets misidentified as photons is estimated by using a control sample of EM-enriched QCD events which are used to calculate the ratio of events that pass the signal photon criteria relative to those that pass really looser criteria in combination with failing the isolation requirement.
Since the EM-enriched sample also includes production of direct single photons, this additional contribution to the ratio is estimated by fitting templates of energy-weighted shower widths from MC-simulated $\gamma+$jets events to an independent QCD data sample, and used to subtract the $\gamma+$jets contribution.
This corrected ratio is applied to a subset of the EM-enriched jet events that passes loose photon identification and additional single-photon event selection criteria, providing a background contribution of $11.2\pm2.8$ jet events.
  
   The 73 observed events in data agree with the total expected background of 75.1 $\pm$ 9.5 events. Distributions in photon p$_{T}$ for the selected candidate events and for those estimated from background are shown in Fig.~\ref{fig:stack_plot}.
The spectra expected from the ADD model for $M_{D}$ = 1 TeV and $n$ = 3 are superimposed for comparison.
Based on these results, exclusion limits are set for the DM and ADD models.

\begin{table}[ht!]
\centering
\caption{Observed (expected) 90\% CL upper limits on the DM model parameters}
{
\begin{tabular}{ccccc}
\hline
$M_{\chi}$ & \multicolumn{2}{c}{Vector} & \multicolumn{2}{c}{Axial-Vector} \\
 & $\sigma$~[fb] & $\Lambda$~[GeV] & $\sigma$~[fb] & $\Lambda$~[GeV] \\
\hline
1           & 14.3 (14.7) & 572 (568) & 14.9 (15.4) & 565 (561) \\
10          & 14.3 (14.7) & 571 (567) & 14.1 (14.5) & 573 (569) \\
100      & 15.4 (15.3) & 558 (558) & 13.9 (14.3) & 554 (550) \\
200      & 14.3 (14.7) & 549 (545) & 14.0 (14.5) & 508 (504) \\
500      & 13.6 (14.0) & 442 (439) & 13.7 (14.1) & 358 (356) \\
1000    & 14.1 (14.5) & 246 (244) & 13.9 (14.3) & 172 (171) \\
\hline
\end{tabular}
\label{tab:LimDM}}
\end{table}

\section{Results and Conclusions}

Upper limits are set on the DM production cross sections, as a function of $M_{\chi}$, assuming vector and axial-vector operators, summarized in Table~\ref{tab:LimDM}.
The obtained limits are converted into the corresponding lower limits on the cutoff scale $\Lambda$, also listed in Table~\ref{tab:LimDM}.
The $\Lambda$ values are then translated into upper limits on the $\chi$-nucleon cross sections, calculated within the effective theory framework.
These cross sections are displayed in Fig~\ref{fig:DMLimits} as a function of $M_{\chi}$ and superimposed with the results of some other experiments.
Previously inaccessible $\chi$ masses below ${\approx}$ 3.5 GeV are excluded for the $\chi$-nucleon cross section greater than ${\approx}$ 3fb at 90$\%$ CL.
For spin-dependent scattering, the upper limits surpass all previous constraints for the mass range of 1--100 GeV.
The results presented are valid for mediator masses larger than the limits on $\Lambda$, assuming unity for the couplings g$_{\chi}$ and g$_q$.

A set of 95$\%$ confidence level (CL) upper limits are also set on the ADD cross sections and translated into exclusions on the parameter space of the model.
The upper limits are calculated using a CL$_{s}$~\cite{CL} method with uncertainties parameterized by log-normal distributions in the fit to data.
The limits on $M_{D}$, with and without $K$-factors, are summarized in Table~\ref{tab:LimADD}.
Masses $M_{D} <$ 1.65 TeV are excluded at $95\%$ CL for $n=3$, assuming NLO cross sections.
These limits, along with existing LO ADD limits from the Tevatron~\cite{CDFgamma,D0gamma} and LEP~\cite{LEPgamma}, are shown in Fig.~\ref{fig:ADDLimits} as a function of $M_{D}$, for  $n=4$ and $n=6$ extra dimensions.
These results extend significantly the limits on the ADD model in the single-photon channel beyond previous measurements at the Tevatron and LEP experiments, and set limits of
 M$_{D}$ $>$ 1.59 -- 1.66 TeV for n=3--6 at 95$\%$ CL.

\begin{table}[ht!]
\centering
\caption{Expected and observed lower limits on $M_{D}$ at 95$\%$ C.L, as a function of extra dimensions n, with K factors(and without, i.e, K = 1)}
{
\begin{tabular}{ccccc}
\hline
n   & {$K$-factors}  & Expected & Observed \\
 &  & $M_{D}$~[TeV] & $M_{D}$~[TeV] \\
\hline
3 & 1.5 & 1.70 (1.53) & 1.73 (1.55) \\
4 & 1.4 & 1.65 (1.53) & 1.67 (1.55) \\
5 & 1.3 & 1.63 (1.54) & 1.64 (1.56) \\
6 & 1.2 & 1.62 (1.55) & 1.64 (1.57) \\
\hline
\end{tabular}
\label{tab:LimADD}}
\end{table}

\end{document}